# Changing our view on design evaluation meetings methodology:

# a study of software technical review meetings

Patrick D'Astous*, Françoise Détienne**, Willemien Visser**

& Pierre N. Robillard*

* École Polytechnique de Montréal, Canada

C.P. 6079, Succ. Centre-Ville

Montreal, Canada, H3C 1A7

dastous@info.polymtl.ca, pierre-n.robillard@polymtl.ca

** EIFFEL - Cognition and Cooperation in Design

INRIA-Rocquencourt

Domaine de Voluceau, BP 105

78153, Le Chesnay

France

Francoise.Detienne@inria.fr, Willemien.Visser@inria.fr

# Changing our view on design evaluation meetings methodology:

# a study of software technical review meetings

## Abstract

By contrast to design meetings, design evaluation meetings (DEMs) have generally been considered as situations in which, according to DEMs methodologies, design activities are quite marginal. In a study of DEMs in software development, i.e. in technical review meetings following a particular review methodology, we showed: (i) the occurrence of design activities as part of an argumentation process; (ii) the relative importance of cognitive synchronisation as a prerequisite for evaluation; (iii) the important role played in evaluation by argumentation that makes explicit the underlying design rationale (DR). On the basis of our results, we discuss the potential for using DR methodologies in this kind of meetings.

Many design methodologies consider design evaluation meetings (DEMs) as places in which, by contrast to design meetings, design activities are supposed to be quite marginal. Often, such activities are even supposed to not occur. Empirical studies conducted on design projects, however, provide elements that lead to believe that design activities do occur in DEMs. Our aim is to identify and to understand the activities actually taking place in this type of meetings.

This paper indeed presents a study of collective cognitive activities taking place during DEMs in software development, i.e. in technical review meetings. Software-development projects involve various types of meetings where team members exchange ideas, evaluate the work that has been done or plan future tasks. These meetings, organised and conducted by software engineers, can be formal or informal. In a typical development process, there are two kinds of meetings: design meetings and design evaluation meetings (DEMs). The objectives of these two types of meetings, as they are prescribed by design methodologies, are quite different, i.e. elaboration of the design for the former vs. evaluation of this design for the latter. Whereas many empirical studies have analysed design meetings, very few of such studies have analysed cognitive and collective activities occurring during DEMs, e.g. inspection meetings or technical review meetings conducted according to particular design methodologies. Existing empirical studies on DEMs are concerned with assessing the effectiveness of various methods by comparing the number of defects detected, rather than to analyse the various activities used in order to detect these defects (see for example, Johnson & Tjahjono[1,2]).

---

[1] **Johnson, P M and Tjahjono D** Assessing software review meetings: a controlled experimental study using *CSRS International Conference of Software Engineering* (1997)

As design methodologies generally consider that design doesn't occur during DEMs, these meetings have not been considered as places to which the use of design rationale (DR) methodologies might be extended. The aim of our studies of software DEMs is to enhance design models, through the account of cognitive and collective activities involved in DEMs, and more generally in collective design evaluation. In addition to this extension of design models, expected application of the results is the development of ergonomic specifications for software-development methodologies, particularly for DEMs: we will consider particularly the role that DR methodologies could play in this kind of meetings.

# *1* STATE OF THE ART

## *1*.1 Design problem characteristics

Our aim is to understand team design activities. From a cognitive point of view, the most common conception of design problems is to consider them as "ill-defined" problems[3,4,5]. Their characteristics are as following:

---

[2] **Johnson, P M and Tjahjono D** Does every inspection really need a meeting? *Empirical Software Engineering*, Vol 3. (1998) pp 9-35

[3] **Simon H** The structure of ill-structured problems *Artificial Intelligence* Vol 4. (1973) pp 181-201

[4] **Visser, W** A Tribute to Simon, and some —too late— questions, by a cognitive ergonomist *Proceedings of the International Conference In Honour of Herbert Simon "The Sciences of Design The Scientific Challenge for the 21st Century" Lyon, Fr (15-16 March 2002)* (edited as INRIA Research Report N° 4462 INRIA, Rocquencourt, Fr)

- the specifications given at the start are never complete or without ambiguity: initial problem specifications are not sufficient to define the goal, i.e., the solution, and progressive definition of new constraints is necessary;

- resolution of conflicting constraints, often coming from different representation and processing systems, plays an important role;

- there is no definite criterion for testing any proposed solution to a design problem, such as there typically exists for "well-defined" problems: various solutions are acceptable, one being more satisfying according to one criterion, another according to another; that is, design problem solutions are not either "correct" or "incorrect,  they are more or less "acceptable";

- problems tend to be large and complex. They are generally not confined to local problems, and the variables and their interrelations are too numerous to be divided into independent sub-systems.

One consequence of this complexity is that the resolution of these problems often requires that multiple competencies be put together, which leads to development of collaboration within a team design group.

With respect to design activity, one may notice several trends in design studies. During a first period, authors of empirical research described design as an activity that followed rather closely the description provided in design methodologies. Later on, however, most researchers started to observe that design

---

[5] **Visser W and Hoc J-M** Expert software design strategies In **J-M Hoc T R G Green R Samurçay and D J Gilmore** (eds) *Psychology of Programming* Academic Press (1990) pp 235-249

was organised opportunistically[6]. With respect to the aspects focused on in this paper, design and evaluation steps were observed to be not taking place sequentially, but to be completely intertwined.

## *1*.2 Design team process

A layered behavioural model of software development processes[7] has been proposed, which underlines the importance of "behavioural", i.e. human and organisational, factors in software productivity. Curtis, Krasner and Iscoe distinguish different layers: individual (with focus on cognition and motivation), team and project (group dynamics), and company and business milieu (organisational behaviour). Our study focuses on the team layer.

Team design can be characterised as cycles of distributed design and co-design phases[8]. In distributed design, each actor has his/her own task to perform. In co-design, several actors have the same common goal. In the distributed design phase, the actors who are simultaneously (but individually) involved in the same co-operation process carry out well-determined tasks. They pursue goals (or at least sub-goals) that are specific to them. In the co-design phase, actors share an identical goal and contribute in order to reach it through their specific skills. They do so with very strong constraints of direct co-operation so as to guarantee a solution to the problem resolution.

---

[6] **Visser W** Organisation of design activities: opportunistic, with hierarchical episodes *Interacting With Computers* Vol 6. N°3 (1994) pp 239-274

[7] **Curtis B Krasner H and Iscoe N** A Field Study Of The Software Design Process For Large Systems *Communications of the ACM* Vol 31. (1988) pp 1268-1287

[8] **Falzon P** Dialogues fonctionnels et activité collective *Le Travail Humain* Vol 57. N° 4/94 (1994) pp 299-312

In individual design, the division of problems into sub-problems is supposed to result in a reduction of complexity —often, however, design problems are difficult to decompose into independent sub-problems, and different decompositions of a same problem are possible[4].

In team design, tasks corresponding to sub-problems are distributed among individuals, each carrying out various sub-tasks. As soon as tasks are divided, disagreement or conflict between designers arises and negotiation generally ensues. Solutions are therefore not only based on purely technical problem-solving criteria. They also result from compromises between designers: solutions are negotiated[9,10,11].

In our present research, focus is on the collective mechanisms involved in co-design. Design methodologies distinguish two kinds of meetings that typically involve co-design: design meetings and design evaluation meetings (DEMs) (e.g., inspection or review meetings). It may be noted that these two

---

[9] **Beguin P** L'activité de travail: facteur d'intégration durant les processus de conception In **P Bossard C Chanchevrier et P Leclair** (eds) *Ingénierie concourante de la technique au social* Economica, Paris, Fr (1997)

[10] **Martin G Détienne F and Lavigne E** Negotiation in collaborative assessment of design solutions: an empirical study on a Concurrent Engineering process *CE'2000, International Conference on Concurrent Engineering* Lyon, Fr (17-20 juillet 2000)

[11] **Martin G Détienne F and Lavigne E** Analysing viewpoints in design through the argumentation process *INTERACT'2001* Tokyo, Jap (July 9-13 2001)

types of meetings could be rather considered as situated on a continuum from design-orientation to evaluation-orientation meetings. The study presented in this paper examines the collective mechanisms involved in co-design through analysis of DEMs conducted according to a particular design methodology prescribing, like other formal technical review methodologies[12],[13], to merely note the presence of defects that are discovered in the software documents under review, and not to determine how to correct them —that is, not to proceed to design of alternative solutions.

Both design and design evaluation meetings group together all or several of the co-designers of a design project, but their objectives as prescribed by design methodologies are quite different. In design meetings, the prescribed objective is to search for a solution to a problem. The accent is on solution development. In DEMs, the prescribed objective is to evaluate and validate the design solutions that are the input to the meeting. The accent is on solution evaluation, and, most of the time, design activities are supposed not to happen —they are even forbidden in certain design methodologies.

Whereas many empirical studies have analysed design meetings, very few of them have analysed activities involved in DEMs, particularly DEMs conducted according to a particular design methodology. Below, we will briefly review the results of the analyses concerning activities involved in these two kinds of meetings, as well as the methodologies developed for these meetings, which correspond to two methodological trends: the Design Rationale (DR) approach for design meetings and, the more general process approach for DEMs.

---

[12] **Fagan M E** Design and Code Inspections to Reduce Errors in Program Development *IBM Systems Journal* Vol 15. N° 3 (1976) pp 182-211

[13] **Fagan M E** (1986) Advances in software inspections *IEEE Transactions on Software Engineering* Vol SE-12. N° 7, pp 744-751

## *1.3 Design meetings*

### *1.3.1 Activities in design meetings*

Previous studies on design meetings have analysed collaboration activities occurring during such meetings: for example, in the development of local area networks[14], of software[15,16,17], of aerospace structures[18], of a mechanical device[19], or of a backpack-to-mountain-bike attachment[20].

---

[14] **Falzon P Darses F and Béguin P** Collective design processes *COOP'96 Second International Conference on the Design Cooperative Systems*, Juan-les-Pins, Fr (June 2-14 1996)

[15] **Herbsleb J D Klein H Olson G M Brunner H Olson J S and Harding J** Object-oriented analysis and design in software project teams In **F Détienne and R Rist** (eds) *Special issue on empirical studies of object-oriented design. Human-Computer Interaction* Vol 10. N° 2 & 3 (1995) pp 249-292

[16] **Walz D B Elam J J Krasner H and Curtis B** A methodology for studying software design teams: an investigation of conflict behaviors in the requirements definition phase In **G M Olson S Sheppard and E Soloway** (eds) *Empirical Studies of Programmers: Second workshop* (pp 83-99) Ablex, Norwood, NJ (1987)

[17] **Olson G M Olson J S Carter M R and Storrosten M** Small Group Design Meetings: An Analysis of Collaboration *Human-Computer Interaction* Vol 7. (1992) pp 347-374

In design meetings, authors have identified, more or less explicitly, different types of activities:

- design activities concerned with the elaboration of solutions and of alternative solutions;

- evaluation activities concerned with the evaluation of solutions or alternative solutions, on the basis of criteria;

- cognitive synchronisation (often referred to as "clarification") activities concerned with the construction of a shared representation of the current state of the solution[8];

- activities pertaining to conflict and negotiation.

Olson et al.[17,21] developed a coding scheme, which contained 22 activity categories depicting the general nature of design discussions. The authors used their coding scheme to analyse the interactions of a group of experienced software designers during design meetings. They found that a great amount of time was being spent on design discussions involving the design themes and their clarification. Herbsleb et al.[15], using a similar coding scheme, compared design meetings in procedural and object-oriented environments. Their goal was to compare the team aspects of these two software development paradigms in order to evaluate the claims made for the superiority of object orientation. They obtained results similar to those of previous studies with respect to the allocation of time in the main categories of activities.

---

[18] **Visser W** Collective design: A cognitive analysis of cooperation in practice In **N F M Roozenburg** (ed) *Proceedings of ICED 93 9th International Conference on Engineering Design* (Vol 1) (pp 385-392) HEURISTA, Zürich, Sw

[19] **Stempfle J and Badke-Schaub P** Thinking in design teams - an analysis of team communication *Design Studies* Vol 23. (2002) pp 473-496

[20] **Cross N Christiaans H and Dorst K** (eds ) *Analysing design activity* (ch 13 pp 271-289) Wiley, Chichester, NJ (1996)

[21] **Olson G M Olson J S Storrotsen M Carter M Herbsleb J and Rueter H** The structure of activity during design meetings In **T P Moran, and J M Carroll** (eds) *Design rationale: concepts, techniques and uses* Erlbaum, Mahwah, NJ (1996)

Stempfle and Badke-Schaub[19] developed a coding scheme distinguishing between six activities related to content issues (search of solution ideas) and five activities related to process issues (management and planning). They found that some teams bypassed cognitive synchronisation (referred to as "analysis") and that this led them to premature evaluation of design ideas. The fact that the observed teams were composed of students in mechanical engineering may partly explain this bias toward premature evaluation.

In a meeting organised in order to design a backpack-to-mountain-bike attachment[20], cognitive synchronisation (referred to as "problem clarifying") is suggested by one of the team members and adopted as the first shared activity of the team. This team member indeed suggests to check first that all participants share the same understanding of the problem rather than to review the existing prototype, as proposed by another team member.

Walz et al.[16] use a particular "process tracing technique", i.e., Bales' "interaction process analysis": group meetings are divided into "interaction process acts" corresponding to an uninterrupted statement (or group of statements) made by one individual, concerning one single issue or context (our "moves", see below). Acts are classified according to a predefined coding scheme, distinguishing three types of acts:

- portrayal of an individual's mental model;

- "bridging" mental models;

- management.

The authors underline the importance of the "issues" discussed, but they neither introduce them explicitly as objects of their "acts", nor distinguish different types of issues.

## *1*.3.2 Methodologies of design meetings

In the traditional approach applied to software design, the process by which the artefacts were produced remained completely implicit. It remained hidden in minutes of meetings, design notebooks or email archives and, mostly, in designers' memories. This way of proceeding made it hard to recover and to reuse design processes. In reaction to this largely "artefact oriented" approach, the design rationale (DR) approach has been created.

In the DR approach, emphasis is on generation and keeping track of design artefacts, both intermediate ones (e.g., requirements and specifications) and the final system. A DR is a representation of the reasoning behind the design of an artefact. Proponents of DR approaches argue that these "rationales" have the potential to play several roles in design[22],[23],[24]:

- structuring design problems;

- maintaining consistency in decision-making;

- keeping track of decisions;

- communicating design rationales to others;

- keeping a chronological record of the design process;

- creating conditions for design reuse.

Several DR notations have been developed to express design reasoning as "arguments" about "issues". Among them, QOC and IBIS are probably the best known. The QOC notation[25] distinguishes Questions, Options and Criteria. The Design Space Analysis approach that uses QOC consists in creating an explicit representation of a structured space of design alternatives and the considerations for choosing among them. It is a process of identifying key problems (Questions), and raising and justifying (via Criteria) design alternatives (Options).

---

[22] **Buckingham Shum S and Hammond N** Argumentation-based design rationale: what use at what cost? *International Journal of Human-Computer Studies* Vol 40. (1994) pp 603-652

[23] **Concklin E J and Burgess K C** A Process-Oriented Approach to Design Rationale *Human-computer Interaction* Vol 6. (1991) pp 357-391

[24] **Moran T P and Carroll J M** *Design rationale: concepts, techniques and uses* Erlbaum, Mahwah, NJ (1996)

[25] **MacLean A Young R M Bellotti V and Moran T** Questions, Options, and Criteria: elements of design space analysis *Human-Computer Interaction* Vol 6. N° 3&4 (1991) pp 201-250

Buckingham et al.[22] have reviewed the empirical evidence of the utility and usability of these techniques in design meetings. Whereas there is evidence that using an argumentative notation augments the reasoning of those who use it, there is also evidence that using such a notation impedes reasoning. The usability claim is not supported by the empirical evidence, which shows that using semiformal schemes for expressing knowledge introduces extra demands.

Furthermore, as emphasised by Buckingham et al., "*the challenge for DR research is to find the most helpful, accessible representations of design reasoning both for initial developers and subsequent designers, which minimise the non-productive effort required to create them* [because capturing useful DR is bound to require some effort]."

One of our aims is to collect data concerning the nature of the activities occurring in design evaluation meetings and their organisation, in order to use these data to specify plausible uses of DRs that "minimise the non-productive effort required to create them". The particular focus of the study presented in this paper are DEMs that follow a particular design methodology according to which design activities are not supposed to occur.

# *1*.4 Design Evaluation Meetings (DEMs)

## *1*.4.1 Activities in evaluation meetings

To date, very few studies have analysed activities involved in meetings specifically dedicated to evaluation, rather then the activities involved in design meetings —among which evaluation plays an important role. As stated before, current empirical studies on DEMs are rather concerned with assessing the effectiveness of various methods by comparing the number of defects detected[1,2]. Some rare studies have examined collaboration activities during evaluation meetings. These were meetings concerned with validation of conceptual database schemes[26] or with code-inspection[27]. The authors have identified activities mainly related to:

- evaluation;

- cognitive synchronisation;

- negotiation.

---

[26] **Karsenty L** *The Validation Dialog for a Database Conceptual Schema* (Research report) INRIA Rocquencourt, France (1991)

[27] **Letovsky S Pinto J Lampert R and Soloway E** A cognitive analysis of code inspection In **G M Olson S Sheppard and E Soloway** (eds) *Empirical Studies of Programmers: Second workshop* Ablex, Norwood, NJ (1987) pp 231-247

These two studies observe, as main difference between design meetings and DEMs, the absence of design activities in the latter kind of meetings. However, both in the study by Karsenty[26] and in the one by Letovsky et al.[27], this absence can be explained by various causes. The evaluation meeting analysed by Karsenty involved both designers and clients. It is likely that the presence of the client inhibits, in some way, design activities by the designers taking place. New requirements are identified in these meetings, but the search for a solution is probably postponed to other meetings that involve only designers. Letovsky et al. analysed a code-inspection meeting. Both the very low level of the artefact being evaluated and the meeting taking place in this implementation phase of the software-development process, probably prevents elaboration of alternative solutions. Implementation revisions are made at the code level, but there is no revision of solutions at a design level.

## *1*.4.2 Methodologies of evaluation meetings

Methodologies for DEMs fall within the perspective of the process model. This kind of model distinguishes different phases occurring in design, from specification to implementation. For each phase, the methodology proceeds to a decomposition into tasks, which may be affected to one or several designers. A DEM may occur after each phase in the global software-development process. Various authors have outlined the activities that are to take place in software DEMs[28],[29]. Design activities are not supposed to occur in such meetings. Solution development is indeed not an objective of DEMs that have two main aims: (1) to verify the current state of the design project, and (2) to validate the specifications for the succeeding tasks.

---

[28] **D'Astous P** *La mesure des activités collaboratives retrouvées lors d'une réunion de révision technique du processus de Génie logiciel* (Rapport EMP/RT-97/30) École Polytechnique, Montréal (1997)

A DEM requires the presence of several reviewers for a certain amount of time (in the order of 30 minutes to two hours, in the meetings reported on in this paper). A project team can hold various DEMs during a typical working week. DEMs serve to facilitate the peer validation of an artefact that is a document representing the actual state of the artefact. This document may be the product of any stage of the design process, from initial requirements to code. Reviewers use existing evaluation criteria.

Each individual plays a role during the meeting. Several authors have identified these roles from a normative viewpoint, that is the one prescribed by the methodology adopted for the DEM[30,31,32,33]. The data collected in the present study has been analysed also, however, from the viewpoint of the actual distribution of activities among the members of the project team according to their roles[34]. In the DEMs observed in this study, the roles in the meetings were organised as follows: a document's "author" (or "producer") managed (as a "moderator") the DEM and ensured the collection of found defects (as a "secretary"). All other participants act as "reviewers".

---

[29] **Johnson P M** *Reengineering Inspection: The Future of Formal Technical Review* (Report ICS-TR-95-24) University of Hawaii at Manoa (1996)

[30] **D'Astous P** Mesure et analyse des activités coopératives lors de réunions de révision technique du processus de génie logiciel (PhD Thesis) École Polytechnique de Montréal, Montréal, Can (1999)

[31] **Humphrey W S** *Managing the Software Process* (ch 10 pp 171-190) Addison-Wesley (1989)

[32] Parnas D L Weiss D M Active Design Review: Principles and Practices *Journal of Systems and Software* Vol 7. (1987) pp 259-265

[33] **Yourdon E** *Structured Walkthroughs* (4th ed) Prentice-Hall (1989)

Due to the prescriptions of DEM methodologies, in particular the interdiction of design activities, DR methodologies didn't find any application in this kind of meetings. As announced above, one of the aims of the study presented in this paper is to analyse the nature of activities actually occurring in design evaluation meetings, in order to examine if DR might be useful also in such meetings.

## *1*.5 *Questions about DEMs*

In previous studies on DEMs, it appeared that design dialogs did not occur. Our hypothesis is that design may occur in such design evaluation meetings. As noticed above, the absence of design dialogs, both in the study by Karsenty[26] and in the one by Letovsky et al.[27], can be explained by various causes: the presence of a client during conceptual database-scheme validation meetings (in the Karsenty study); the very low level of the artefact in code-inspection meetings (in the Letovsky et al. study).

A DEM is an evaluation meeting that may take place at any phase in the development process. Furthermore, all participants of these meetings are designers. Thus, this kind of evaluation meeting is quite different from the two evaluation meetings referred to before. In fact, the characteristics of this kind of meetings make them quite similar to design meetings: the same kind of participants (all designers), the temporal occurrence (in any phase, that is, all during the design process).

---

[34] **D'Astous P Détienne F Robillard P N and Visser W** Quantitative measurements of the influence of participants roles during peer review meetings *Empirical Sotware Engineering* Vol 6. (2001) pp 143-159

On the basis of these similarities, it is likely that the same kind of design activities as in design meetings will occur (even if, according to DEM methodologies, they are not supposed to take place). More particularly, we suppose that such design activities, if indeed they occur in DEMs, will rather take place in the context of argumentation, e.g., in order to convince one's peers, than in that of design development as such.

So, the study presented in this paper aims to enhance existing design models, especially with respect to cognitive and collective activities involved in DEMs, and more generally in collective design evaluation meetings. Its specific objectives are, resuming:

- to collect data concerning the nature of the activities occurring in a DEM and their organisation (e.g., through argumentative movements), specifically in DEMs conducted according to a methodology prescribing to merely note the presence of defects in the software documents under review;

- to formulate, on the basis of these data, ergonomic specifications for software-development methods, particularly for design evaluation meetings, especially focusing on the relevance of methods for keeping track of the reasoning underlying the design of an artefact (such as DR).

## *2 METHODOLOGY*

In our empirical study, we observed the activities of team members during DEMs in a software development project.

## *2. 1 Data*

### *2.1.1 Characteristics of observed technical review meetings*

All DEMs that were studied took place during one and the same particular software development project. The goal of this project was to develop a business process simulator based on Petri Nets. The project involved four full-time software engineers and lasted for 19 weeks. The project team used a defined and documented software engineering process estimated to be at level CMM-2, i.e. level 2 of the Capability Maturity Model for software[35].

The DEMs studied, that is technical review meetings, were mandatory meetings required before each documented activity. They were being held after each detailed design stage of the development process (Figure 1). During these DEMs the objective was to merely note the presence of defects that were discovered in the software documents under review, and not to determine how to correct them —that is, not to proceed to design of alternative solutions. The acceptance of the document at the end of the meeting was a necessary condition for going on to the next step, i.e. coding (in the implementation stage). The resulting design documents were composed of verbal, natural language descriptions, algorithms and formal object-oriented notations. The paradigm used throughout the development process was the object-oriented paradigm.

---

[35] **Paulk M C Curtis B Chrissis M B and Weber C V** *Capability Maturity Model for Software, version 1 1* (Report SE-93-TR-24) Software Engineering Institute (1993)

## 2.1.2 Software project team

The same four team-members participated in all observed DEMs (all male). A particular team member who was the author of the document to be reviewed, distributed his document a day or so in advance so that everyone could read the document before the start of the DEM. The pace of the meeting was dictated by the time spent on the various sections of the document. A meeting ended on a decision as to the general acceptability of the document.

## 2.1.3 Data collection

The observational approach used is to videotape the DEMs, because this does not disturb the meetings and thus allows valid data to be collected. After a time, people forget that they are being videotaped and they begin to use familiar expressions and very technical terms, talk at the same time, interrupt and tell jokes; in short, exhibit very "natural" behaviour.

A specially trained typist transcribes each video into a **protocol**. A good transcript is not trivial to make and various trials were needed before the right way to do so was found. The protocols form the basic data for the analysis, although it is sometimes useful to refer back to the video to validate the meaning of a statement, e.g. through a speaker's intonation.

Seven meetings (DEMs) from the software development project have been videotaped, transcribed into protocols and analysed for the present study. Each meeting protocol is cut up into individual participant verbal **turns**, according to change of speaker.

## 2.2 Data coding and segmentation method and Analysis

### 2.2.1 Theoretical basis

Several coding and segmentation methods based on cognitive ergonomics[15] and conversational linguistics[36] have been reviewed. In conversational linguistics, four units are generally considered pertinent in the description of the organisation of conversations[37]. These are moves, exchanges, sequences and (verbal) interactions (see Figure 2).

A verbal **interaction** is a communication unit that shows evident internal continuity, but breaks with what precedes or follows it, in terms of time, space and participants, and of subjects (themes) discussed. Working meetings, such as DEMs, are a specific type of verbal interactions. It is therefore possible to borrow the concepts developed for their representation in the study of DEMs. The moves, exchanges and sequences of seven interactions (DEMs) are studied in this paper.

A verbal interaction can be divided into verbal **sequences** according to discussed subjects. The software design solution that is the main theme of a DEM, and is represented in the document under review, is subdivided into sub-solutions corresponding to distinct parts of the document. The document's linear structure guides the occurrence of themes during a typical DEM. This situation was typical of our observed DEMs: this may be influenced either by the document templates or the object-oriented

---

[36] **Kerbrat-Orecchioni C** *Les Interactions Verbales* (tome 1) Armand Colin, Paris, Fr (1990)

[37] **Roulet E Auchlin A Moeschler J Rubattel C and Schelling M** *L'articulation du discours en français contemporain* Peter Lang (1985)

paradigm used during the development. Each DEM is therefore divided into verbal sequences according to the parts of the document discussed. Other themes can be derived from the original theme during a verbal sequence. These nested themes are still part of the original verbal sequence.

A sequence is composed of one or more **exchanges**, each one characterised by one functional activity rather than by one theme. Examples of collective functional design activities are solution elaboration and evaluation.

A **move** is the contribution of a single speaker to a given sequence and is characterised by one **activity** and the **subject** (theme) that is the object of the activity. This unit may correspond to, or be a subset of, an individual verbal turn[38].

In linguistics, it is usual to distinguish levels of moves, reflecting embeddednes of themes introduced in a sequence. Whereas a sequence identifies all the moves relating to the same sub-solution under discussion, sub-discussions may take place reflecting exchanges on other introduced subjects, like criteria or hypotheses formulated by participants. This possibility is particularly important for the analysis of discursive inter-moves linking.

Several empirical studies have been carried out on design meetings through observations conducted on the activities of the participants. In our opinion, none was however completely appropriate to represent the earlier mentioned units used to describe polylogues (dialogues conducted between more than two

---

[38] **Sinclair A Coulthard R M** *Towards an Analysis of Discourse* Oxford University Press, Oxford, UK (1975)

participants). Furthermore, there are DEM activities that are not accounted for in previous studies on design meetings.

Olson's coding method[17,21] as well as the one used by Stempfle and Badke-Schaub[19] do not distinguish activities from subjects (or themes). A meeting is therefore represented as a flat structure in these analyses. Segmentation of meetings into sequences according to themes and derived, nested themes is impossible with these methods, making further analyses less accurate.

In order to translate the movement between levels of abstraction in the discussion of issues, Walz et al.[16] represent the issues addressed by the group during a meeting as nodes in a hierarchical issue chart for this meeting, by reference to an abstraction hierarchy (general - specific) (cf. QOC). But, as noted above, these issues are not linked to the acts. Concerning such acts, Karsenty's coding method[26] has a higher level of granularity, using exchanges (called "dialogues" in his terminology) as the lowest level of analysis. Karsenty does not take verbal moves into account.

## 2.2.2 Representation of moves

### 2.2.2.1 Coding scheme

Five characteristics are needed to describe a specific move, facilitate its identification and provide the basis for coherent analysis:

- ID: identifies the move through the speaker and the rank in the polylogue;

- TYPE: identifies a move as a request or an assertion;

- ACTIVITY: identifies the speaker's action;

- SUBJECT: identifies the informational entity on which the activity is performed;

- ATTRIBUTE: complements the subject, generally with respect to a criterion (criteria can concern either form or content, see Table 3).

The ACTIVITY and SUBJECT characteristics form the core of the move descriptions.

The definition of these five characteristics was obtained through an iterative process of experimentation and validation of the coding scheme on one typical meeting. Experts in cognitive ergonomics and software engineering experts completed validation.

An elementary syntax, composed of the separator symbol / between each characteristic, is used to code a move in such a way that it will facilitate automatic analysis. The generic code is then:

ID/TYPE/ACTIVITY/SUBJECT/ATTRIBUTE

The default value of TYPE is "assertion"; only the "request" type will appear in the coding.

For example, in the 60$^{th}$ move by teammate M (ID), M makes an assertion (TYPE) consisting in the evaluation (ACTIVITY) of the design solution x (SUBJECT), referring to the format used to describe x (ATTRIBUTE). Its code is:

M60/EVAL/SOLx/CRIT.F

We identified different types of activities, subjects and criteria (see Table 1).

## *2*.2.2.2 Activities

Table 2 presents a list of the activities with their definitions.

## *2*.2.2.3 Subjects

An activity is always performed on a subject. A subject is the link that maintains cohesion in a discussion. Each move either introduces a new subject of discussion or elaborates on a previously introduced subject. People do not introduce subjects randomly, but rather in relationship to the ongoing discussion. It is usual that a given move may become the subject of a forthcoming move ("Result of a previous activity" in Table 1). A statement like "I agree with what you said…" is a typical example where the subject of the activity (Acceptance) is made up of the previous move (what has been said). Each move persists in time since a participant may react to someone else's move right away, after a certain time, or maybe never. People remember what has been said.

## *2*.2.2.4 Attributes

The design document (that is, the artefact under review) is divided into sections that become the subjects of *Introduction* activities. Criteria (Table 3) are attributes used to further characterise a given move. There are two types of criteria: form and content criteria. Form criteria are used when participants discuss the format of a subject, rather than its technical relevance. They are often based on software engineering guidelines. The content criteria used in the meetings observed in this study come from standard ISO 9126[39].

Criteria are coded by CRIT.C (content attributes) and CRIT.F (form attributes).

## 2.2.2.5 Example of a protocol coded with respect to moves

In order to examine the reliability of our coding scheme, in particular the inter-rater reliability, one of the seven DEMs was coded by two persons, one of this paper's authors and a Master student of the same department. The degree of accordance between the two coders[40] was quite good, as measured by two statistical tests: the kappa of Cohen (0.75) and the index of fiability of Perrault and Leigh (0.79).

Because of the time needed to code the whole protocol, protocol coding was afterwards realised by only one person, one of this paper's authors. This author was familiar with the application domain and more particularly with the observed project, but was not a member of the project team. Figure 3 shows a sample coding extracted from the analysis of a DEM. Codes such as "SOLed" and "SOLee" refer to solutions corresponding to sections of the design document under review. The criteria (CRIT.C and CRIT.F) are differentiated by lowercase letters.

---

[39] **ISO** *Information Technology- Software quality characteristics and sub-characteristics* (ISO/IEC 9126-1) (1991)

[40] **Labelle S** *Validation d'une étude empirique qualitative en génie logiciel* (Mémoire de maîtrise) Université de Montréal, Montréal, Can (1999)

Some explanations on the coding may be useful. Move 51 is an implicit introduction of SOLed into the discussion. An implicit introduction is systematically added to the protocol when a new subject is discussed without being explicitly introduced by a speaker. In move 61, *a priori* the terms "Ah, Ah" may have many meanings (acceptance, request, etc.). At this point, a look back at the video is necessary to get a feel of the intonation. In this case, it was clearly an acceptance of the previous move. Moves 62, 63 and 64 are subsets of one verbal turn where the speaker first introduced a solution (62), justified its content (63), and then further described the solution by giving information not contained in the original document (64). The subject of moves 63 and 64 (INTRO62) is in fact the result of a previous move (62).

## *2.2.3 Representation of sequences*

A sequence is a grouping of related moves whose main subject is one and the same part of the document (Solx). There may be nested subjects derived from discussions occurring earlier in the sequence. The scope of a sequence is defined by the relationships between the moves. A primary move, which often introduces a subject, is the basis for a sequence. A sequence is composed of this primary move and all the succeeding moves that elaborate on the subject of the primary move or on other previous moves within the sequence. A sequence may contain moves at several levels. Two rules are used to differentiate these levels:

- Rule A: A move is nested with respect to a reference level if the subject of its activity belongs to the set of objects instantiated by any move at the reference level. Attributes have no influence within this rule.

- Rule B: A move is nested with respect to a reference level if its subject is an attribute used at the reference level.

Figure 4 shows an example of sequences coding. The first sequence refers to SOLed, which was introduced implicitly in move 51, producing subject INTRO51. In moves 52 and 54, B and M review the functionality of SOLed. Speaker C then hypothesises on the nature of SOLed, which creates the object HYP57. Then M rejects this hypothesis based on an attribute of content (CRIT.Ca), which creates the object REJ60. Finally C accepts the rejection (REJ60) created by move M60. Then a second sequence starts on the basis of the new subject SOLee.

## *2*.2.4 Representation of exchanges

A verbal exchange is part of a sequence, composed of one or several moves, and characterised by one particular functional activity rather than by a particular subject. Examples of collective functional design activities are solution elaboration and evaluation.

Each exchange is defined through the different composing activities and its subjects. The " | " means "or"; (**ACTIVITY**) refers to an optional activity:

**Cognitive Synchronisation (Synch)**: This exchange enables the participants to make sure that they share a common representation of the state of (alternative) design solutions or of evaluation criteria (content or form). It is characterised by:

**INFORM | HYPOTHESISE**

**Subjects**: solutions, criteria or alternative developments

**(ACCEPT)**

**Subjects**: information or hypothesis

**Review (Rev)**: This exchange enables participants to evaluate the value of, or to give their opinion on, (alternative) design solutions or criteria used in the solution review. An evaluation can either be negative or positive. Results of a review, especially positive ones, may remain implicit. This exchange is characterised by:

**EVALUATE | JUSTIFY**

**Subjects**: solutions, criteria or alternative developments

**(ACCEPT)**

**Subjects**: evaluation or justification

**Conflict Resolution (Conf)**: This exchange enables argumentation between two or more participants regarding a conflict generated by diverging opinions on criteria or (alternative) solutions, or by diverging representations of the state of the (alternative) design solutions or of evaluation criteria. It is characterised by:

**REJECT**

**Subjects**: solutions, criteria, alternative developments, or hypotheses

**(ACCEPT)**

**Subject**: rejection

**Alternative Elaboration (Alt. Elab.)**: This exchange enables design and analysis of solutions not present in the document being revised. It is characterised by:

**DEVELOP**

**Subjects**: alternative developments or solutions

**(ACCEPT)**

**Subject**: alternative development

**Management (Man)**: This exchange enables co-ordination and planning of different tasks:

**MANAGE**

**Subjects**: tasks (meetings or projects)

These patterns have been derived using an empirical approach requiring experts in cognitive ergonomics in order to study the categories of activities characterising the meetings and to derive exchange components on the basis of activity patterns (Figure 5). We have validated some of these patterns using a statistical approach, Lag Sequential Analysis (LSA), which enables the identification of categories of activities that follow one another (see §3.2).

## 3 RESULTS

Seven DEMs from the software development project have been analysed. They were composed of 148 sequences (each made up of a certain number of participant verbal moves). We didn't take into consideration the 21 sequences that were related to project management. The results presented in this paper were obtained from the analysis of the remaining 127 sequences: their subjects were parts of the document describing the solution to be reviewed. In what follows, we will not distinguish the data from the seven DEMs because a similar distribution of moves characterised the different meetings.

## *3.1 Distribution of moves over levels*

Figure 6 shows the distribution of moves and words per level. First, we establish that both measures are correlated. Given this result, we will further only present distribution of moves within each level. Second, we found that most moves are located at level 1 and level 2, representing together 88% of moves, while levels 3 and more embedded levels represent only 8% of the moves.

Level 0 includes all introductions of artefacts (subjects) in the discussion, while level 1 includes all moves whose subjects are these introduced artefacts only. Approximately 67% (level 0 and 1) of all moves are directly related to the discussion of the reviewed artefact. These moves represent 74% of words that compose the sequences.

Given the distribution per level, we have focused our further analyses only on level 1 and level 2.

## *3.1.1 Level 1*

Figure 7 shows the distribution of activities at level 1. By definition, the different sub-solutions composing the solution corresponding to the design project are the subjects of all the moves at the first level. One observes that, at this level, although the exchanges are very focused on the solutions, the main activity is not reviewing. We observed that the most frequent exchanges were, in this order, cognitive synchronisation (41%, corresponding to hypothesis, HYP, and information, INFO, activities), review (38%, justification, JUS, and evaluation, EVAL) and alternative elaboration (21%, development, DEV).

Two remarks may be formulated on the basis of these results. The main objective of DEMs is to review a document that represents a state of the software design project. However, we found that most time of the discussion on the artefact under review is not spent in reviewing activities (38% of the time), but rather in cognitive-synchronisation activities (41%). We also found that the team spends nearly as much time in alternative-elaboration activities (21%) —which are typical of design— as in evaluation activities (26%), even though, according to the software methodology used by the observed team, design should not take place in a DEM.

## 3.1.2 Level 2

Nesting level 2 contains all the subjects previously generated (at level 1). Different moves can therefore have different subjects as opposed to the unique subject at level 1. Figure 8 shows that acceptance and justification activities are predominant ( 46% and 18% respectively).

An analysis of level 2 subjects relates activities to particular subjects (see Figure 9). In fact, a third of the moves (34%) are related to alternative solutions that have been generated. Noteworthy is the appearance of criteria as subjects of level 2 moves (11%).

The nature of the activities related to these subjects is, however, quite divers (Figure 10). The fact that criteria are taken as subjects of exchanges at nested levels indicates that there is cognitive synchronisation (INFO and HYP, 83%) and review (JUS and EVAL, 17%) of the DEM procedure by the participants. For example, the participants explain the nature of a particular criterion for solution evaluation or they evaluate the criteria by order of priority. This means that part of the functional activities of the team is to make explicit and to evaluate the evaluation procedure. This is not a task that is prescribed in the DEM methodology.

Alternative solutions are being reviewed (78%, EVAL, JUS and ACC) for their adequacy. These results suggest, indirectly, that alternative solutions are being generated during a DEM, even though these solutions are not included in the documents that store the results of the meeting. The generation of a new solution is usually followed by activities where participants attempt to judge its nature (evaluation).

## *3.2 Configuration of exchanges*

On the basis of frequent configurations of exchanges, one can qualify those which are characteristic of evaluation meetings. In order to identify such typical configurations of exchanges (i.e. configurations occurring with a significant frequency), instances of two types of methods have been used, i.e. two quantitative statistical methods and one qualitative grammatical method. The statistical methods used are Lag Sequential Analysis[41] (LSA) and hierarchical clustering[42]; the grammatical method applies rewriting rules[43].

The analysis of sequential structures using LSA is grounded in information theory. LSA enables the identification of units (moves, exchanges) that follow each other, with or without other units in-between. The analysis consists in determining whether or not the occurrence frequency of a given unit is independent of the occurrence frequency of another unit. Sequential structures enable the definition of configurations. This quantitative approach can validate groupings identified by a qualitative analysis (such as the exchanges presented earlier).

These methods have been used in an iterative way, in several cycles, as long as LSA allows significant configurations to be detected. One or more rewriting rules are applied on the results of each cycle.

Rewriting rules are used in order to group into new units occurrences of configurations in a structure. Hierarchical clustering is used in order to examine similarities inside sequential structures. Both are utilised as long as the LSA applied on the new resulting structure provides significant outcomes (i.e. exchanges or sequences) (for more details see Robillard, D'Astous, Détienne & Visser[44]).

---

[41] **Allison P D Liker J K** Analyzing Sequential Categorical Data on Dyadic Interaction: A comment on Gottman *Psychology Bulletin* Vol 91 (1982) pp 393-403

[42] **Johnson S C** Hierarchical Clustering Schemes *Psychometrika* Vol 32. (1967) pp 241-254

[43] **Gonzalez R C Thomason M G** *Syntactic Pattern Recognition: An Introduction* Addison-Wesley (1978)

In our analysis of DEMs (see §3.1), we found that the most frequent activities were cognitive synchronisation and review, followed by alternative elaboration. This led us to suppose that there is a tight connection between cognitive synchronisation and review activities (translating evaluation which one might suppose to be typical of DEMs) on the one hand, and between review and design activities (supposed to be implemented, mainly, by elaboration of —alternative— solutions) on the other hand. Through the present analysis in terms of exchanges and sequences, we have examined the nature of these supposed relationships.

After a certain number of cycles applying LSA and rewriting, followed by a hierarchical-clustering step, the following links between exchanges have been identified as significant at $p < .05$ (see Figure 11).

Figure 11 may be read as follows. When introduction of a solution (INTRO) is followed immediately by a development (DEV), this development consists in changing the form of the solution, and there is an implicit negative evaluation according to a criterion of form. Introduction of a solution can also be followed immediately by either its evaluation alone (EVAL) or its evaluation and development of an alternative solution (in one order or another, i.e. DEV-EVAL or EVAL-DEV). Such review activities may, or may not, be preceded by a cognitive-synchronisation exchange (HYP-INFO or INFO-HYP). In these latter cases, the evaluation bears mainly upon content criteria.

---

[44] **Robillard P N D'Astous P Détienne F and Visser W** Measuring cognitive activities in software engineering *International Conference of Software Engineering (ICSE98)* Kyoto, Jap (April 19-25, 1998)

With respect to the relationship between review and cognitive synchronisation: when review is introduced by cognitive synchronisation, this means that a shared representation of the to-be-evaluated subject may be a prerequisite for its review to take place. The argumentative movement is of the type "proposition-opinion".

With respect to the relationship between review and design: the review of a solution, in particular a negative review, leads participants to make explicit alternative solutions: the solution may be a justification for the negative review or an alternative solution for the currently rejected solution. The argumentative movement is of the type "opinion-arguments".

N.B. On the surface, the arguments may be presented, either before the opinion they support is presented (DEV-EVAL), or afterwards (EVAL-DEV).

On the basis of the configuration of exchanges, we see that the activities of elaboration and of cognitive synchronisation, even if not expected in the DEM prescribed task, are both necessary and useful in collaboration taking place through argumentation.

## 3.3 Design rationale graph

An analysis of several review sequences has identified the existence of underlying design rationale in participant discussions. Most review sequences will start with a Question (explicit or implicit) concerning the original Option presented (i.e., part of the document under review). This Question will, in turn, engender different Options from the participants. These Options are approved or rejected according to existing Criteria, or to new Criteria presented during the sequence. Figure 12 shows two rationales: one concerning the content of a solution and one concerning the form of a solution. In the case of the form design rationale, it is clear that the programming guide, advocated by the procedure expert, has a lot of strength in the final decision.

Formal capture of this information may prove beneficial to the development project as a whole. A design rationale could even replace the classic meeting minutes that do not really reflect the actual nature of the discussions, as highlighted in a complementary analysis[45].

## 4. DISCUSSION AND CONCLUSION

This paper aimed at answering two questions: (1) Which are the activities that actually take place in DEMs? and (2) Is it possible to identify the Design Rationale from DEMs? The analyses performed provided the necessary data to answer these questions.

Our analysis of DEMs shows that the actual activities in the DEM exchanges do not correspond to the prescribed task of the team. The DEM methodology prescribes solution evaluation, whereas it prohibits design. Designers participating in a DEM are supposed to merely note the presence of defects that they discover in the software documents they are reviewing: they are not to determine how to correct these defects —that is, they are not to proceed to design of alternative solutions. We observed, however, exchanges that are typical of collective design. Thus, evaluation meetings cannot do without design of alternative solutions —even if methodologies or other prescriptive guidelines forbid such activities. Evaluation during DEMs even prompts design activity! Such design exchanges are largely part of an argumentation process used to justify the negative evaluation of a solution. It thus is an untenable principle of design methodologies to instruct design reviewers not to engage in the development of (new) design alternatives during DEMs —or other (formal) technical review meetings!

---

[45] **Ipperciel S** *Caractérisation des réunions de révision technique dans un projet de génie logiciel* (Master thesis) École Polytechnique Montréal, Montréal, Can (1998)

Another result of the present study is the relative importance of cognitive-synchronisation exchanges as compared to review exchanges, which are supposed to be the main objective of DEMs. Relationships between evaluation and clarification activities have already been shown in Karsenty[26]. The present study shows that a shared representation of the to-be-reviewed subject is a prerequisite for evaluation activities to occur: evaluation is indeed often introduced by cognitive synchronisation.

Given this observation of cognitive synchronisation as a prerequisite for review, and given the duration of all synchronisation exchanges in each meeting, one may imagine that the institution of an explicit synchronisation phase at the beginning of each meeting or each sequence (i.e. discussion of a part of the solution implemented in the document under review) may, on the one hand, spare time, and on the other hand assure that the co-designers have a shared understanding of the subject under discussion, and not only of those elements which, by chance, have been the subject of synchronisation. This will, of course, not stop some clarification activities from occurring during a meeting —if only because requests about what needs to be clarified may also arise during interaction.

We also showed that cognitive synchronisation concerns not only the solution to be evaluated, but also the review procedure. In this case, participants make explicit their criteria and order them, which reveals construction of common knowledge, required in context, on the evaluation procedure. The same kind of results has been found also by Darses[46], who observes that evaluation of solutions leads to making criteria explicit. This result is in accordance with Curtis et al.[7] who observed that the process of a team coming to "common representational conventions" could take, in early phases of development, as much time as did the use of the conventions themselves.

Argumentation plays an important role in evaluation. Our study shows that the underlying design rationale (DR) is made explicit through this argumentation process. A DR method such as QOC could be used to report the evaluation of the reviewed artefact. With respect to the notions of Questions, Options and Criteria, their identification has been proven not to be easy in design meetings, especially in upstream design where the space of possible problems and solutions is still very open. In DEMs, it is likely that their identification will be easier because:

- previous tasks, for most of them individual design tasks, have already ensured partial structuring of the problem and solution spaces;

- due to the collective nature of the activity during DEMs, argumentation plays an important role in these meetings and thus making DR explicit should be perceived as an intrinsic activity rather than as an added task.

Adding a DR to each reviewed artefact would increase the amount of available information for future design and DEMs in the software process. This could be accomplished by integrating DR much more tightly with the artefact under development. In DEMs, reviewed artefacts may remain as documentation of the design project. DR methodology could facilitate the review of this artefact and could be used to extend the documentation, being integrated with the artefact itself.

---

[46] **Darses F** L'ingénierie concourante: Un modèle en meilleure adéquation avec les processus cognitifs en conception In **P Bossard C Chanchevrier et P Leclair** (eds) *Ingénierie concourante de la technique au social* Economica, Paris, Fr (1997)

The use of DRs in DEMs is thus consistent with some recent directions of research for DR[23].

- Using DR as a management tool in the process-oriented approach of DR. In this way it would be a productive effort with immediate feedback. This could be relevant with regard to the double objective of DEMs, i.e., to verify the current state of the design project and to validate the specifications for the succeeding tasks.

- Limiting the practice of DR to specific events such as meetings, be it design meetings or design evaluation meetings, e.g. technical reviews, in order to make them easier to accept.

  - If DR is used both in design and evaluation meetings, evaluation could take advantage of the rationale produced during design.

- Integrating DR much more tightly with the artefact under development. The artefact being "constructed" may be any product of the design process, from initial requirements to code (cf. the trend called "construction driven argumentation" by Fischer et al.[47]).

---

[47] **Fischer G Lemke A C McCall R and Morch A I** Making argumentation serve design *Human-Comptuter Interaction* Vol 6. N° 3 & 4 (1991) pp 393-419

- It could be interesting to consider this integration as an annotation technique of the artefact, i.e., a secondary level of information attached to a main document, which has a special status in the project.

- In DEMs, the reviewed artefact will remain as documentation of the design project. DR methodology could facilitate the review of this artefact and could be used to extend the documentation, being integrated with the artefact itself. A design rationale could even replace the classic meeting minutes, which do not really reflect the actual nature of the discussions.

Generalisation of our results must be limited by the fact that they are based on the analysis of DEMs held in the specific detailed-design phase of only one software development project. Future work will consist in the analysis of other project environments in order to support the generalisation of these results. Furthermore, the use of DR in the further evolution of software development methodologies will be investigated in the future.

## *ACKNOWLEDGEMENTS*

**Table 1 Review move activities, subjects and attributes**

| Activities | Manage (project or meeting) (MAN) |
| --- | --- |
| | Introduce or read part of the document to be reviewed (INTRO) |
| | Develop (DEV) |
| | Evaluate (EVAL) |
| | Formulate a hypothesis (concerning the "what?", "how?" or "why?") (HYP) |
| | Inform ("what?" or "how?") (INFO) |
| | Justify ("why?") (JUS) |
| | Accept (ACC) |
| | Reject (REJ) |
| **Subjects** | Solution (implemented in part of the document to be reviewed) (SOLx) |
| | Project or meeting (exclusively subject of MAN) (PROJ and MEET) |
| | Result of a previous activity (coded by the activity and the sequence ID) |
| **Attributes** | Form criterion (CRIT.F) |
| | Content criterion (CRIT.C) |

## Table 2 Reviewing activities

| ACTIVITY | ABREV. | DEFINITION |
|---|---|---|
| Management | MAN | Coordinating and planning the different tasks at the project or meeting level. |
| Introduction | INTRO | Introducing a new subject into the discussion. |
| Development | DEV | Presenting a new idea in detail. |
| Evaluation | EVAL | Judging the value of a subject. An evaluation is negative, positive or neutral. |
| Hypothesis | HYP | Expressing a personal representation of a subject, using phrases such as "I believe that…", "I think …" or "Maybe…". |
| Information | INFO | Handing out new knowledge with respect to the nature of a subject. |
| Justification | JUSTIF | Arguing or explaining the rationale of a choice. |
| Acceptance | ACC | Considering a subject as being valid. |
| Rejection | REJ | Discarding a subject as being invalid. |

**Table 3 List of attributes**

| Criteria of form | Criteria of content (ISO 9126) |
|---|---|
| Nomenclature | Functionality |
| Algorithms | Reusability |
| Documentation | Portability |
| Functions | Reliability |
| Files | Maintainability |
| Data Types | Efficiency |
| Editor | Ease of implementation |
| Variable declaration | |
| Global variables | |
| Document structure | |
| Semantics | |
| Level of description | |

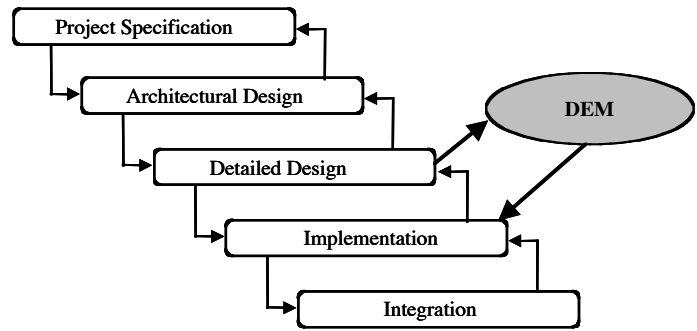

**Figure 1 Observed software process**

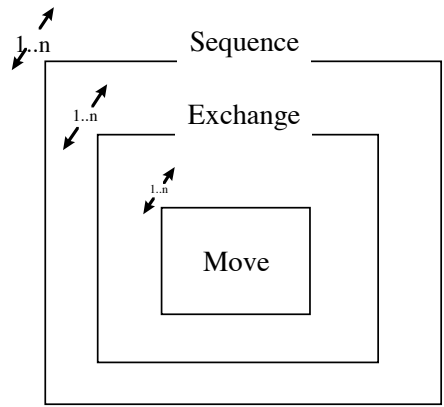

Verbal interaction

**Figure 2 Units of description**

|   | ID | Move | Coded move |
|---|----|------|------------|
|   | 51 |      | --51/INTRO/SOLed |
| B | 52 | Why did you put 150 there? | B52/REQ/JUSTIF/INTRO51 |
| M | 54 | I don't believe in using 150 DEFINE. These will do the same thing, but the compiler will check them while the compiler doesn't check DEFINEs. | M54/JUSTIF/INTRO51/CRIT.Ca |
| C | 57 | There may be more than 50 error messages, you know! | C57/HYP/INTRO51 |
| M | 60 | Ah no, this is JUSTIF a type, like the type of the message itself. | M60/REJ/HYP57/CRIT.Ca |
| C | 61 | Ah, Ah. | C61/ACC/REJ60 |
| M | 62 | It's JUSTIF that I need, I need some fields OK, these four fields there! | M62/INTRO/SOLee |
| M | 63 | Because I need some fixed arrays at the start for the messages. | M63/JUSTIF/INTRO62/CRIT.Ca |
| M | 64 | So, I fix them, I fix the first four. The additional messages will follow. We'll be able to put whatever we want, an error message, insufficient memory. | M64/INFO/INTRO62/CRIT.Ca |
| B | 65 | Why then if we can use them any way! | B65/INFO/INTRO62 |
| B | 67 | Yeah, OK, we don't have a choice. | B67/ACC/INTRO62/CRIT.Ca |
| M | 68 | We can do it here. | M68/INFO/INTRO62 |

**Figure 3 Example of a protocol coded with respect to moves**

Note. Inaudible statements and moves concerning "regulators", e.g. moves expressing that a participant is listening to the current speaker(s), have been eliminated.

| | ID | Move | Coded move | Level | Rule | Sequence |
|---|---|---|---|---|---|---|
| | 51 | | --51/INTRO/SOLed | 0 | | |
| B | 52 | Why did you put 150 there? | B52/REQ/JUSTIF/INTRO51 | 1 | A | |
| M | 54 | I don't believe in using 150 DEFINE. These will do the same thing, but the compiler will check them while the compiler doesn't check DEFINEs. | M54/JUSTIF/INTRO51/CRIT.Ca | 1 | A | |
| C | 57 | There may be more than 50 error messages, you know! | C57/HYP/INTRO51 | 1 | A | Sequence 1 |
| M | 60 | Ah no, this is JUSTIF a type, like the type of the message itself. | M60/REJ/HYP57/CRIT.Ca | 2 | A | |
| C | 61 | Ah, Ah. | C61/ACC/REJ60 | 3 | A | |
| M | 62 | It's JUSTIF that I need, I need some fields OK, these four fields there! | M62/INTRO/SOLee | 0 | A | |
| M | 63 | Because I need some fixed arrays at the start for the messages. | M63/JUSTIF/INTRO62/CRIT.Ca | 1 | A | |
| M | 64 | So, I fix them, I fix the first four. The additional messages will follow. We'll be able to put whatever we want, an error message, insufficient memory. | M64/INFO/INTRO62/CRIT.Ca | 1 | A | Sequence 2 |
| B | 65 | Why then if we can use them any way! | B65/INFO/INTRO62 | 1 | A | |
| B | 67 | Yeah, OK, we don't have a choice. | B67/ACC/INTRO62/CRIT.Ca | 1 | A | |
| M | 68 | We can do it here. | M68/INFO/INTRO62 | 1 | A | |

Figure 4 Identification of sequences and level of moves

| ID | | Move | Coded move | Exchange | Level | Rule | Sequence |
|---|---|---|---|---|---|---|---|
| | 51 | | --51/INTRO/SOLed | | 0 | | |
| B | 52 | Why did you put 150 there? | B52/REQ/JUSTIF/INTRO51 | REVIEW | 1 | A | |
| M | 54 | I don't believe in using 150 DEFINE. These will do the same thing, but the compiler will check them while the compiler doesn't check DEFINEs. | M54/JUSTIF/INTRO51/CRIT.Ca | | 1 | A | |
| C | 57 | There may be more than 50 error messages, you know! | C57/HYP/INTRO51 | COGN. SYNCHRO. | 1 | A | Sequence 1 |
| M | 60 | Ah no, this is JUSTIF a type, like the type of the message itself. | M60/REJ/HYP57/CRIT.Ca | CONFLICT | 2 | A | |
| C | 61 | Ah, Ah. | C61/ACC/REJ60 | | 3 | A | |
| M | 62 | It's JUSTIF that I need, I need some fields OK, these four fields there! | M62/INTRO/SOLee | | 0 | A | |
| M | 63 | Because I need some fixed arrays at the start for the messages. | M63/JUSTIF/INTRO62/CRIT.Ca | REVIEW | 1 | A | |
| M | 64 | So, I fix them, I fix the first four. The additional messages will follow. We'll be able to put whatever we want, an error message, insufficient memory. | M64/INFO/INTRO62/CRIT.Ca | COGN. SYNCHRO. | 1 | A | Sequence 2 |
| B | 65 | Why then if we can use them any way! | B65/INFO/INTRO62 | | 1 | A | |
| B | 67 | Yeah, OK, we don't have a choice. | B67/ACC/INTRO62/CRIT.Ca | | 1 | A | |
| M | 68 | We can do it here. | M68/INFO/INTRO62 | | 1 | A | |

**Figure 5 Sample coding of exchanges**

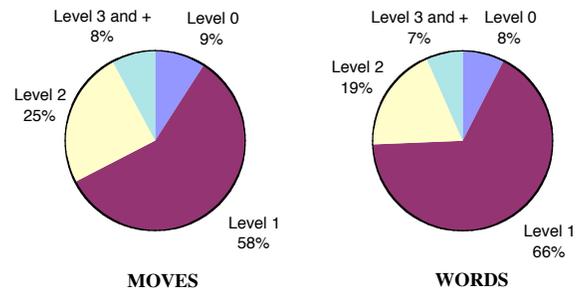

**Figure 6 Move distribution per level**

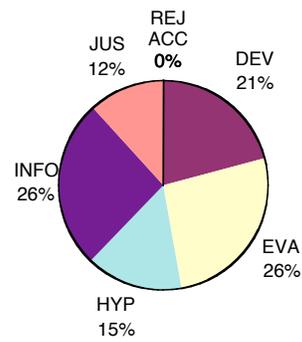

**Figure 7 Relative distributions of activities at level 1**

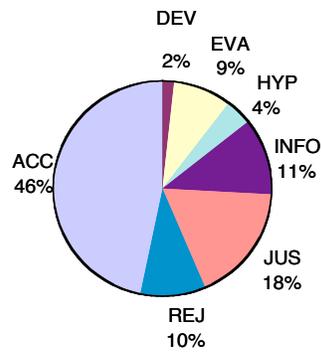

**Figure 8 Level 2 activities relative distribution**

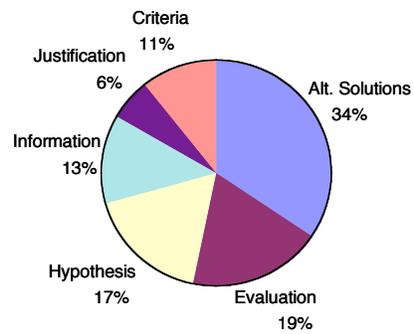

**Figure 9 Level 2 subjects relative distribution**

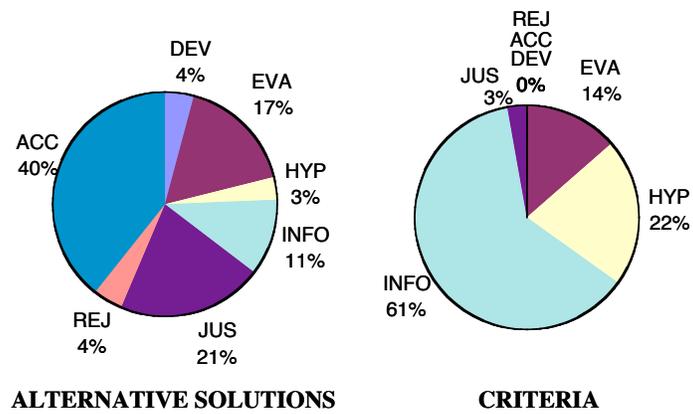

**Figure 10 Relative distributions of activities at level 2**

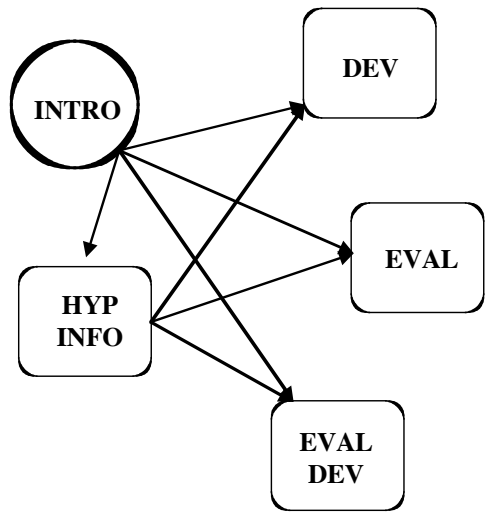

**Figure 11. Sequential structures in DEMs**

Key

Represents a move

Represents an exchange

Represents a significant transition

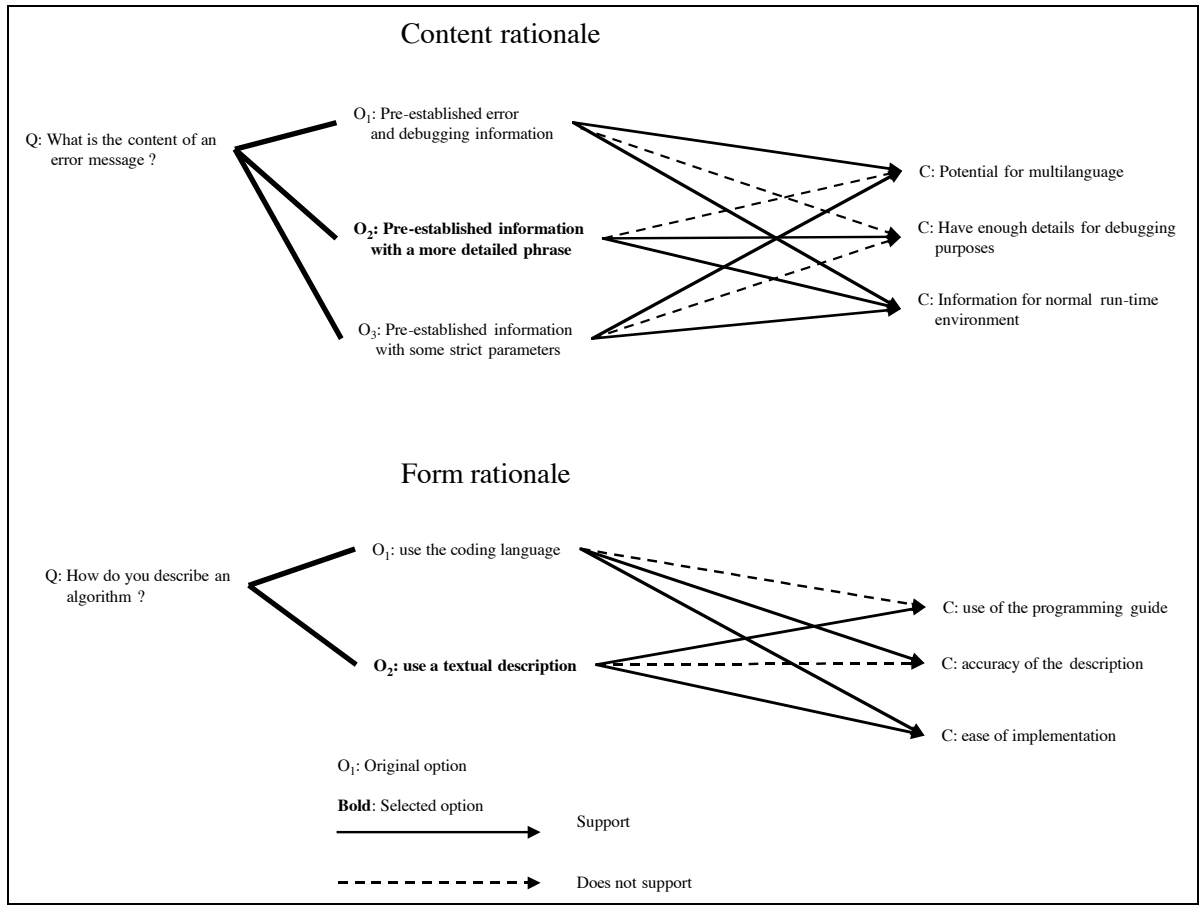

**Figure 12 Content and Form Design Rationale**